# High-energy nitrogen rings stabilized by superatom properties

Zhen Gong [a, #], Rui Wang [a, #], Famin Yu [a, #], Chenxi Wan [a], Xinrui Yang [a] and Zhigang Wang *[, a, b]

[a] Institute of Atomic and Molecular Physics, Jilin University, Changchun 130012, China

[b] International Center for Computational Method & Software, College of Physics, Jilin University, Changchun 130012, China.

[#] The authors contribute equally to this work.

* Author to whom correspondence should be addressed: wangzg@jlu.edu.cn

## Abstract

How to stabilize nitrogen-rich high-energy-density molecules under conventional conditions is particularly important for the energy storage and conversion of such systems and has attracted extensive attention. In this work, our theoretical study showed for the first time that the stabilization mechanism of the nitrogen ring conformed to the superatomic properties at the atomic level. This result occurred because the stabilized anionic nitrogen rings generally showed planar high symmetry and the injected electrons occupied the superatomic molecular orbitals (SAMOs) of the nitrogen rings. According to these results, we identified the typical stabilized anionic nitrogen ring structures $N_6^{4-}$, $N_5^-$ and $N_4^{2-}$, and their superatomic electronic configurations were $1S^21P^41D^41F^22S^21P^21F^21D^42P^41G^41F^4$, $1S^21P^41D^41P^22S^21F^41D^42P^4$ and $1S^21P^41D^21P^21D^22S^22P^41D^4$, respectively. On this basis, we further designed a pathway to stabilize nitrogen rings by introducing metal atoms as electron donors to form neutral $ThN_6$, $LiN_5$ and $MgN_4$ structures, thereby replacing the anionization of systems. Our study highlights the importance of developing nitrogen-rich energetic materials from the perspective of superatoms.

## Introduction

Nitrogen-rich molecules are considered good candidates for high-energy-density materials (HEDMs) since they can release a large amount of energy and their decomposition products are clean and nonpolluting [1-8]. At present, many nitrogen-rich molecules have been synthesized, such as $K_{72}(N_6)_4(N_5)_{56}(N_2)_{72}$, $(N_5)_6(H_3O)_3(NH_4)_4Cl$ and $FeN_4$ crystals, in which the most important energetic component is the planar nitrogen rings $N_6$, $N_5$ and $N_4$ with high symmetry [3-5]. Interestingly, the structural features are very consistent with one species of superatom—planar superatom [9, 10]. Although the perspective connecting the energetic nitrogen rings and superatoms has not been established, this shows a promising analytical perspective for determining the stability of these high-energy nitrogen rings.

Superatom, as a special molecule due to its stability and wide varieties, has attracted significant attention in the fields of physics, chemistry, and material science [11-19]. Synthesized clusters, such as $Na_n$ and $Al_{13}^-$, are considered to be superatoms [14, 15], and their stability mechanism relies on the superatomic properties of atom-like electron configurations[14, 20, 21]. Additionally, the stability of fullerenes $C_{60}$ and $B_{40}$, which can exist in natural environments, has been demonstrated to be related to these superatomic properties [22-25]. Interestingly, superatoms also contain special planar species, such as benzene [26] and its derivatives and coinage metal planar structures [27-29], and they satisfy superatomic electronic configurations with two-dimensional ductility. All these results indicate that although the discovery process of superatomic properties may lag behind the preparation process of corresponding cluster molecules, superatoms are increasingly accepted. It also provides a promising direction for achieving the stability of cluster molecules.

In this work, we report our study on the stability mechanism of the recognized nitrogen rings at the atomic level. The typical nitrogen rings can be successfully stabilized to form anionic structures $N_6^{4-}$, $N_5^-$, and $N_4^{2-}$ by injecting electrons. More importantly, these anionic nitrogen rings clearly exhibit the electronic structural properties of planar superatoms with high symmetry. Based on this, a principle method for forming stable nitrogen rings was developed, which involved selecting suitable metal atoms as electron donors to interact with the nitrogen rings. Our study provides an effective direction for the design and regulation of nitrogenous HEDMs.

## Methods

In this work, the studied system was fully optimized using third-generation dispersion-corrected density functional theory. The hybrid generalized gradient approximation exchange-correlation functional B3LYP was used since it had the advantages of computational efficiency as well as accurately predicting the structure of small clusters and calculating correlation energies of electron densities [30]. In the calculation of geometric configurations, the (14s13p10d8f6g)/[10s9p5d4f3g] valence basis set (RECP small nucleus approximation) for the Th atom was used [31], and the 6-311++G** [32] basis set for other atoms was used. This method was also used for natural population analysis (NPA) [33]. To ensure that the structures were energy-minimized on the potential energy surface, the vibrational frequencies were calculated to verify the geometric structures at the same theoretical level. All above calculations used the Gaussian09 program [34]. The Multiwfn 3.8 package [35] was used to analyze the electronic structure, Mayer bond orders, adaptive natural density partitioning (AdNDP) [36]and electron localization function (ELF) [37].

## Results and discussion

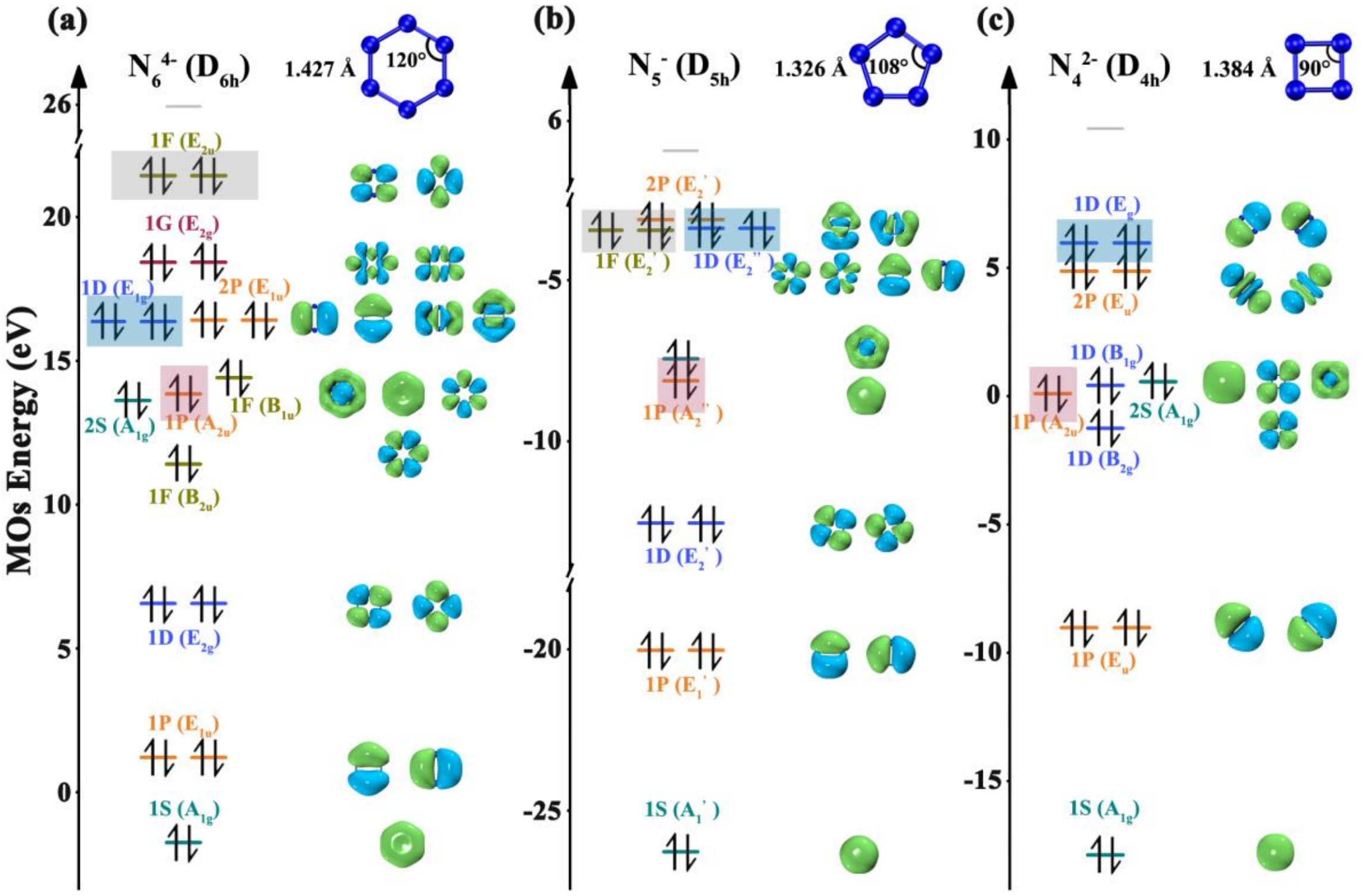

**Figure 1.** Molecular orbital energy levels and geometric structures of $N_6^{4-}$, $N_5^-$ and $N_4^{2-}$. Superatomic electronic configurations: a, $1S^21P^41D^41F^22S^21P^21F^21D^42P^41G^41F^4$ for $N_6^{4-}$. b, $1S^21P^41D^41P^22S^21F^41D^42P^4$ for $N_5^-$. c, $1S^21P^41D^21P^21D^22S^22P^41D^4$ for $N_4^{2-}$. The geometric structure of the nitrogen rings is shown in the top right position. For the displayed MO diagram, an isosurface of 0.02 is used.

Through screening and optimizing different nitrogen rings with different charge states, stable structures were finally obtained, which contained anionic nitrogen rings $N_6^{4-}$, $N_5^-$ and $N_4^{2-}$ (structural coordinate see Table S1 of Supporting Information (SI)). As shown in Fig. 1, their N–N bond lengths were 1.427 Å ($N_6^{4-}$), 1.326 Å ($N_5^-$) and 1.384 Å ($N_4^{2-}$), and their N-N-N bond angles are 120°, 108° and 90°, with symmetries of $D_{6h}$, $D_{5h}$ and $D_{4h}$, respectively. These geometric parameters indicated that these anionic nitrogen rings formed high-symmetry planar structures.

Next, the molecular orbital (MO) energy levels of these anionic nitrogen rings were further examined from the perspective of superatomic electron structures. Analogizing atomic states in a single atom with a well-defined shell structure (s, p, d, f, etc.), the angular momentum number of MOs was identified by inspecting the global as we did for atomic orbitals. These MOs were classified by the irreducible representation of the point-group symmetry [38]. In Fig. 1(a), starting from the lowest energy, the MO ($A_{1g}$) resembled the s atomic orbitals and was identified as 1S SAMO. The next two groups of doubly degenerate MOs ($E_{1u}$ and $E_{2g}$), which resembled the p and d atomic orbitals, were identified as 1P and 1D SAMOs, respectively. The next nondegenerate MO ($B_{2u}$) resembled the f atomic orbitals and was identified as 1F SAMOs. Similar to the above method, the three nondegenerate MOs ($A_{1g}$, $A_{2u}$ and $B_{1u}$) and two groups of doubly degenerate MOs ($E_{1g}$ and $E_{1u}$) were sequentially named 2S, 1P, 1F, 1D and 2P SAMOs. The next doubly degenerate MOs ($E_{2g}$) resembled the g atomic orbitals and were identified as 1G SAMOs. The final doubly degenerate MOs ($E_{2u}$) were identified as 1F SAMOs. Based on the above results, the superatomic electronic configuration of $N_6^{4-}$ was $1S^21P^41D^41F^22S^21P^21F^21D^42P^41G^41F^4$. Based on the same discussion as above for $N_5^-$ and $N_4^{2-}$, the MOs of $N_5^-$ were $A_1'$ (1S), $E_1'$ (1P), $E_2'$ (1D), $A_2''$ (1P), $A_1'$ (2S), $E_2'$ (1F), $E_1''$ (1D), and $E_1'$ (2P), and the MOs of $N_4^{2-}$ were $A_{1g}$ (1S), $E_u$ (1P), $B_{2g}$ (1D), $A_{2u}$ (1P), $B_{1g}$ (1D), $A_{1g}$ (2S), $E_u$ (2P), and $E_g$ (1D) in increasing order of energy. Therefore, their superatomic electronic configurations were $1S^21P^41D^41P^22S^21F^41D^42P^4$ and $1S^21P^41D^21P^21D^22S^22P^41D^4$, respectively. Their corresponding SAMO shapes are displayed on the right side of each graph in Fig. 1. Regarding the occupied MOs, the result indicated that the additional electrons occupied the SAMOs and kept the systems from reducing symmetry. Therefore, their electron configurations enabled them to possess the near-planar superatomic properties [27, 39]. This mechanism was the primary reason behind the stability of the anionic nitrogen rings [21, 24-27, 40]. To further demonstrate that anionic nitrogen rings with superatomic properties are more stable than corresponding neutral rings, we conducted binding energy and bond order analyses [41]. The results showed that $N_4^{2-}$, $N_5^-$, and $N_6^{4-}$ had higher binding energies (-5.34, -7.22, and -5.96 eV per atom) than their corresponding neutral nitrogen rings (-4.49, -6.08, and -1.64 eV per atom). Additionally, the bond orders of the weakest N-N bonds in $N_4^{2-}$, $N_5^-$, and $N_6^{4-}$ (1.74, 1.78 and 1.74) were higher than those in their corresponding neutral nitrogen rings (1.60, 1.71, and 1.70) (for details see Table S2 and S3 in the SI). These results validated the higher stability of nitrogen rings with superatomic properties.

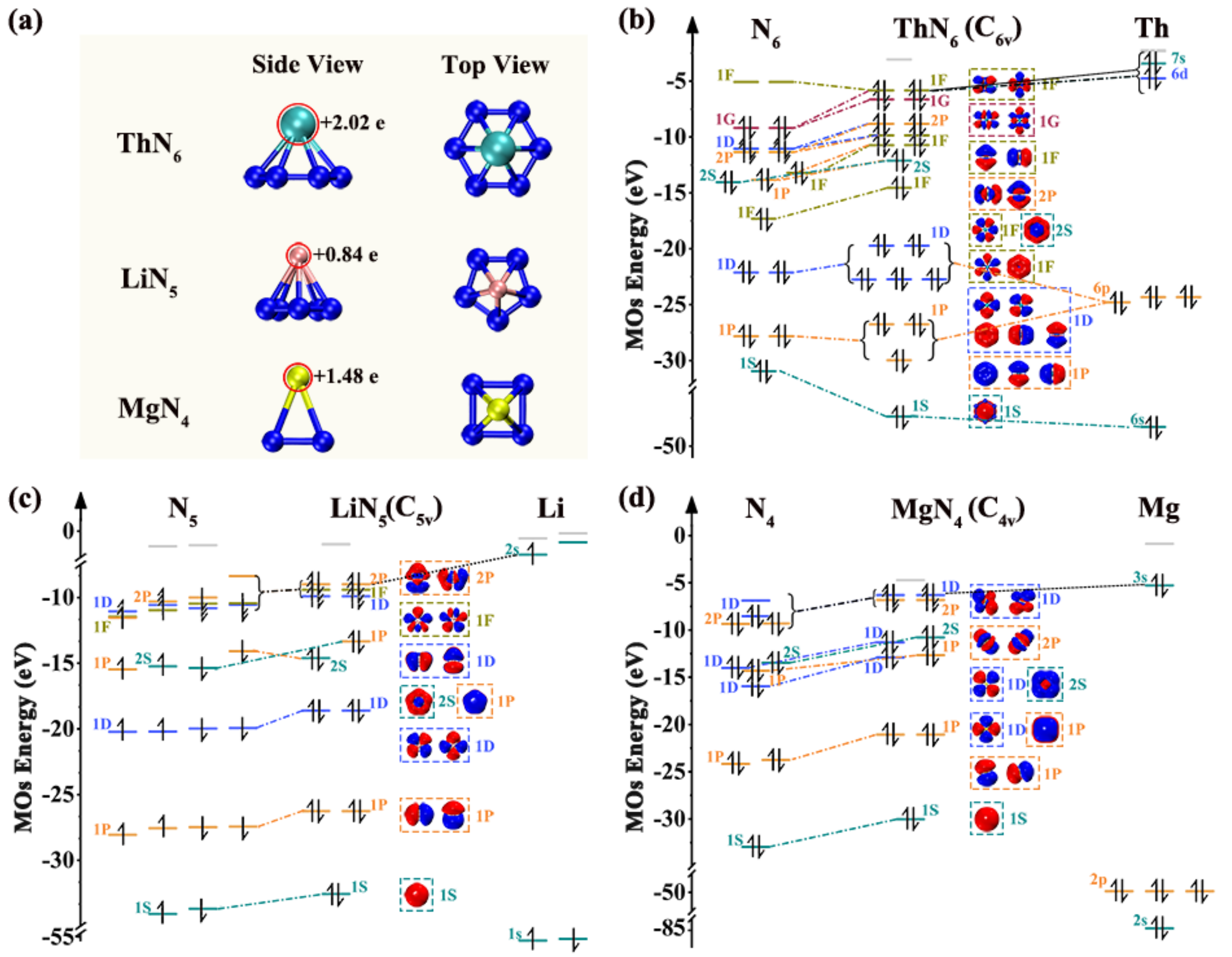

**Figure 2.** Structure and energy level diagrams of $ThN_6$, $LiN_5$ and $MgN_4$. a, Structure diagrams of $ThN_6$, $LiN_5$ and $MgN_4$. The charges show the amount of positive charges transferred from the metal atom to the nitrogen rings by NPA. b, c and d, Energy level diagrams of SAMOs in $ThN_6$, $LiN_5$ and $MgN_4$. The corresponding shapes of the 1S, 1P, 1D, and 1F SAMOs are shown on the right side of the diagrams. For the displayed MO diagrams, an isosurface of 0.02 is used. The short dashed dotted lines represent shared electrons, and the dotted lines represent charge transfer.

Based on the anionic structure of these nitrogen rings and that the metal atoms can be used as electron donors to stabilize fullerene-type structures [42-44] due to their active valence electrons, we further examined the feasibility of introducing different metal elements as electron donors to stabilize the $N_6$, $N_5$ and $N_4$ systems. Through numerous attempts, the metal-stable nitrogen rings of $ThN_6$, $LiN_5$ and $MgN_4$ were formed when Th, Li and Mg metal elements were the electron donors interacting with $N_6$, $N_5$ and $N_4$, respectively (structural coordinate see Tables S4 of SI); from Fig. 2(a), the symmetries of these metal-stable nitrogen rings were $C_{6v}$, $C_{5v}$ and $C_{4v}$ respectively, and their ground states were all singlets. Specifically, in each system, the nitrogen ring maintained the symmetry of an isolated anionic nitrogen ring. This result demonstrated that it was effective to achieve stabilization of the nitrogen rings with this approach.

To understand the interactions of metals and nitrogen rings, charge decomposition analysis (CDA) was performed on the systems ($ThN_6$, $LiN_5$ and $MgN_4$), where every system was divided into two fragments: metals and nitrogen rings; the energy level diagrams of SAMOs are shown in Fig. 2(b)~(d). The SAMOs of $ThN_6$ had a similar shape to that of the $N_6^{4-}$ system, and these SAMOs were mainly caused by the fragment $N_6$. In particular, the highest occupied orbitals (HOMO) and adjacent HOMO-1 of $ThN_6$ were caused by the lowest unoccupied orbitals (LUMO) and adjacent LUMO+1 of $N_6$, as well as the valence orbitals on 6d and 7s of the Th atom. This result indicated that the Th atom not only acted as an electron donor but also shared certain electrons with $N_6$, which enabled the system to maintain the superatomic electronic configuration, namely, $1S^21P^61D^{10}1F^22S^21F^82P^41G^41F^4$. For $LiN_5$ and $MgN_4$, their SAMOs were mostly contributed by the SAMOs of the nitrogen rings. In particular, the HOMO of metal-stable nitrogen rings was mostly contributed by the LUMO of the nitrogen rings. Similarly, the metal atoms Li and Mg also played the role of electron donors, which enabled $LiN_5$ and $MgN_4$ to maintain their superatomic electronic configurations of $1S^21P^41D^41P^22S^21D^41F^42P^4$ and $1S^21P^41D^21P^21D^22S^21D^42P^4$, respectively.

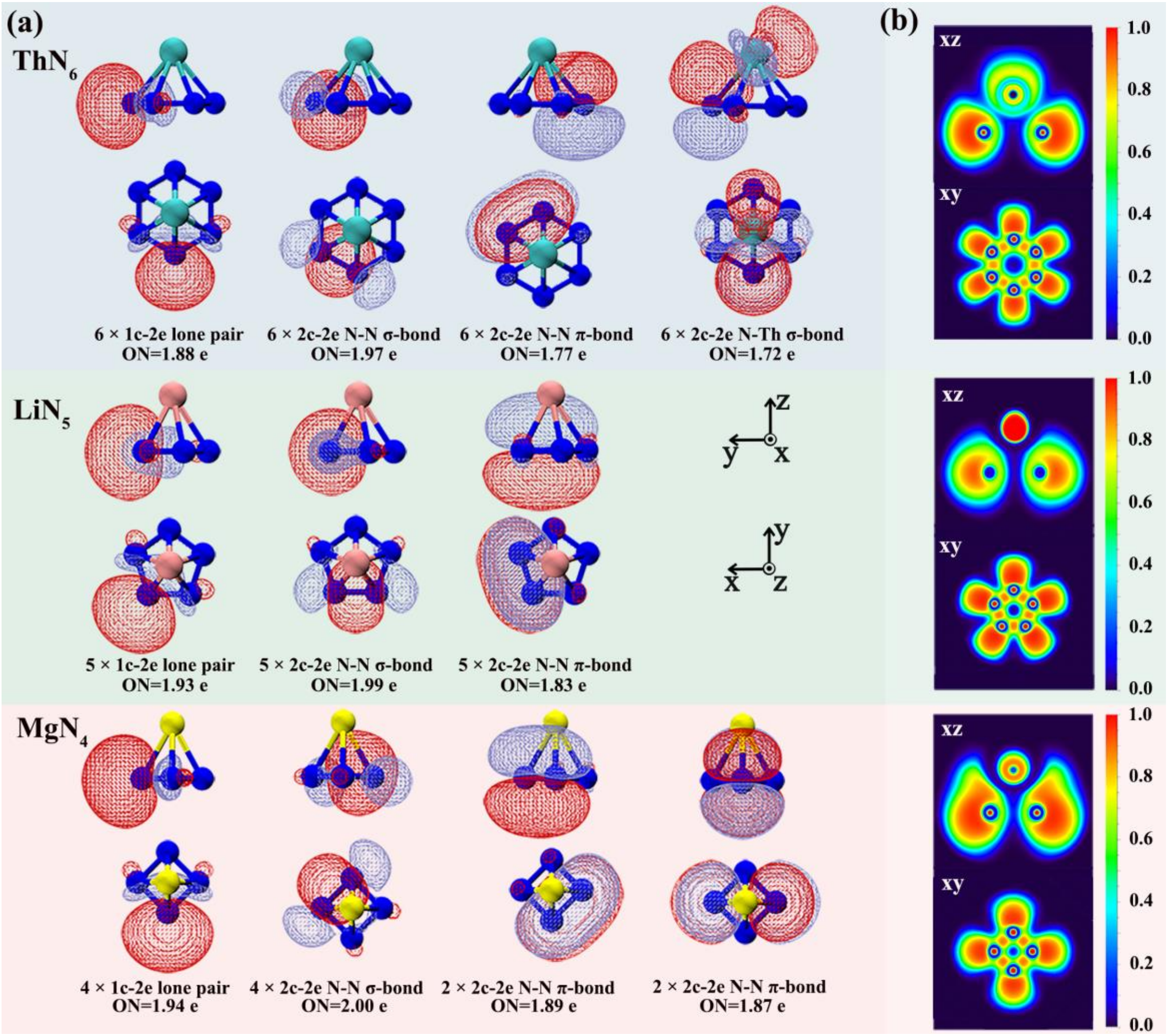

**Figure 3.** Bonding patterns between metal atoms and nitrogen rings. a, AdNDP bonding patterns of $ThN_6$, $LiN_5$ and $MgN_4$. The $b \times a$-center two-electrons ($a$c-2e) means that the number of $a$c-2e is b. The ON represents the number of electrons to form $a$c-2e. Covalent bonding patterns between two N atoms (N–N) contain π-bonds and σ-bonds, and σ-bonds are only formed between N and Th atoms (N-Th). b, ELF maps of $ThN_6$, $LiN_5$ and $MgN_4$ on the xy and xz planes.

To further explore the interactions between metal atoms and nitrogen rings, natural population analysis (NPA) [33] was performed. In Fig. 2(a), the negative charges transferred from metal atoms to nitrogen rings were 2.02 e ($ThN_6$), 0.84 e ($LiN_5$) and 1.48 e ($MgN_4$), and these charges were equally distributed to each nitrogen (N) atom of nitrogen rings (for details see Table S5 of SI). This result indicated that the nitrogen rings became an anionic structure, which was consistent with the results of CDA. Specifically, the 1F SAMOs of $ThN_6$ indicated that there were more than two charges transferred between the Th atom and nitrogen ring $N_6$, which differed from the result of NPA. Thus, the interactions between metal atoms and nitrogen rings were complex. Therefore, it was necessary to further analyze the bonding patterns between the metal atoms and nitrogen rings to determine the type of interaction between them.

Adaptive natural density partitioning (AdNDP)[36] was performed to systematically analyze the bonding pattern between metal atoms and nitrogen rings. As an effective method for extending natural bond orbital (NBO) analysis [45], AdNDP showed bonding in terms of $a$-center two electrons (ac-2e) [46]. In our work, we primarily focused on 1c-2e and 2c-2e, which represented lone-pair electrons of atoms and covalent bonds between two atoms, respectively. Moreover, the occupied number (ON) represented the number of electrons to form $a$c-2e, and the ON was provided for those for which ON was over 1.70 (see Fig. 3(a) and Fig. S1 of SI). In $ThN_6$, there were six 1c-2e N lone-pair electrons (ON=1.88 e), six 2c-2e N–N σ-bonds (ON=1.97 e), six 2c-2e π-bonds (ON=1.77 e) and six 2c-2e N-Th σ-bonds (ON=1.72 e). Thus, approximately 11.28 valence electrons were distributed around N atoms, approximately 22.44 valence electrons were distributed to the N–N junctions, and approximately 10.62 valence electrons were distributed at the N-Th junctions. The detailed analysis of the remaining valence electrons after forming 2c-2e bonds showed that there were approximately -0.29 valence electrons on each N atom and 2.12 valence electrons on the Th atom. This result indicated that the Th atom transferred approximately two electrons to $N_6$, which was in agreement with the results of the charge transfer analysis. The distribution of the remaining valence electrons after forming N-Th σ-bonds (2c-2e) showed that there were 0.11 valence electrons on the Th atom. Thus, the covalent bond between Th and N atoms mainly originated from the two valence electrons of the Th atom. Therefore, the interaction between Th and the nitrogen ring was caused by both ionic interactions and covalent interactions.

In addition, in $LiN_5$, there were five 1c-2e N lone-pair electrons (ON=1.93 e), five 2c-2e N–N σ-bonds (ON=1.99 e) and five 2c-2e N–N π-bonds (ON=1.83 e) (see Fig. S2 of SI). In MgN4, there were four 1c-2e N lone-pair electrons (ON=1.94 e), four 2c-2e N–N σ-bonds and four 2c-2e π-bonds (ON=2.00 e, 1.87~1.89 e) (see Fig. S3 of SI). Similar to $ThN_6$, the corresponding analysis indicated that one electron was transferred from Li to $N_5$, and two electrons were transferred from Mg to $N_4$. Therefore, ionic interactions occurred between the metal atoms and nitrogen rings.

To further characterize covalent interactions between metal atoms and nitrogen rings, the electron localization function (ELF) [37], which exhibited the degree of electronic delocalization between fragments, was analyzed. The numbers 0 and 1 represented no localization (blue) and full localization (red), respectively. In Fig. 3(b), for $ThN_6$, there was a covalent interaction between them. However, for $LiN_5$ and $MgN_4$, the color between the Li atom and the nitrogen ring $N_5$, as well as between the Mg atom and the nitrogen ring $N_4$, was blue, which indicated that there was no covalent interaction between them. This result was also confirmed by Fozzy bond order [41] analysis, as shown in Table S4 of the SI, and was consistent with the AdNDP analysis. Particularly, there were covalent and ionic interactions between the Th atom and nitrogen ring $N_6$; however, there was no covalent interaction between the Li atom and nitrogen ring $N_5$ or between the Mg atom and nitrogen ring $N_4$.

## Conclusions

In summary, the stabilization mechanism of anionic nitrogen rings included the formation of planar structures with high symmetry and satisfied superatomic properties by introducing electrons. Based on this mechanism, we further developed a method to stabilize nitrogen rings by introducing metals as electron donors. In this case, the nitrogen rings in the formed systems still conformed to the superatomic properties. We anticipated that this work will facilitate new insight into the stabilization of high-energy nitrogen rings, expand the understanding of their fundamental properties, and provide an important theoretical basis for the design and synthesis of novel energy-containing superatomic materials.

## Author contributions

Z. Gong and R. Wang and F. Yu performed simulations and calculations. Z. Wang initiated and supervised the work. Z. Gong, R. Wang, F. Yu, C. Wan, X. Yang and Z. Wang analyzed the results. Z. Gong, R. Wang, F. Yu and Z. Wang contributed to writing the paper.

## Conflicts of interest

All authors declare no competing interests.

## Acknowledgements

The authors wish to acknowledge Ms. Yue Xin and Ms. Hongbo Jing, for discussion. This work was supported by the National Natural Science Foundation of China (grant numbers 11974136 and 11674123). Z. Wang also acknowledges the assistance of the High-Performance Computing Center of Jilin University and National Supercomputing Center in Shanghai.